# Scavenging of oxygen vacancies at modulation-doped oxide interfaces: Evidence from oxygen isotope tracing

Y. Z. Chen[1], M. Döbeli[4], E. Pomjakushina[2], Y. L. Gan[1], N. Pryds[1], and T. Lippert[2,3]

[1]Department of Energy Conversion and Storage, Technical University of Denmark, Risø Campus, Roskilde 4000, Denmark
[2]Paul Scherrer Institute, Research with Neutrons and Muons Division, CH-5232 Villigen PSI, Switzerland
[3]Department of Chemistry and Applied Biosciences, Laboratory of Inorganic Chemistry, ETH Zürich, CH-8093 Zürich, Switzerland
[4]Ion Beam Physics, ETH Zurich, CH-8093 Zürich, Switzerland

**Abtract:**

The introduction of manganite buffer layers, La$_{7/8}$Sr$_{1/8}$MnO$_3$ (LSMO) in particular, at the metallic interface between SrTiO$_3$ (STO) and another band insulator suppresses the carrier density of the interfacial two-dimensional electron gas (2DEG) and improves significantly the electron mobility. However, the mechanisms underlying the extreme mobility enhancement remain elusive. Herein, we used $^{18}$O isotope exchanged SrTi$^{18}$O$_3$ as substrates to create 2DEG at room temperature with and without the LSMO buffer layer. By mapping the oxygen profile across the interface between STO$^{18}$ and disordered LaAlO$_3$ or yttria-stabilized zirconia (YSZ), we provide unambiguous evidence that redox reactions occur at oxide interfaces even grown at room temperature. Moreover, the manganite buffer layer not only suppresses the carrier density but also strongly suppresses the oxygen exchange dynamics of the STO substrate, which likely prevents the reduction of STO during the formation of the 2DEG. The underlying mechanism on the enhanced electron mobility at buffered oxide interfaces is also discussed.

Strontium titanate, SrTiO$_3$ (STO), is a representative perovskite-type oxide insulator with a band gap of 3.2 eV. Analogous to silicon for conventional semiconductor electronics, STO is the basis material for oxide electronics. It is by far one of the most widely used substrate materials for the growth of epitaxial oxide thin films due to its structural compatibility to both isostructural perovskites and non-isostructural spinels, fluorites, pyrochlores, and so on. Doped STO with oxygen vacancies or with substitutional elements such as La, Nb or Ta, in its own right, shows a wide range of interesting properties such as metallic conduction with high electron mobility [1] and superconductivity [2]. Besides, STO is also a model system for mixed ionic and electronic conductors at high temperatures, due to the presence of unintentional dopants (typically Fe, Al, and Mn) which act as acceptor type impurities [3]. In recent years, STO has attracted increasing attention for application in oxide electronics as active materials, particularly upon the discovery of a two-dimensional electron gas (2DEG) at the STO surface [4] or its interface with another band insulator, such as perovskite LaAlO$_3$ (LAO) [5] or spinel γ-Al$_2$O$_3$ (GAO) [6].

The 2DEG of STO interface to another oxide insulator has drawn extensive attention because of the emergent properties which are not observed in the bulk counterparts, such as 2D superconductivity [7,8], magnetism [9], quantum Hall effect [10], as well as light enhanced field effects [11]. On the other hand, although the 2DEG of bare STO surface is widely accepted to originate from the formation of surface-confined oxygen vacancies and concomitant band bending [4], the driving force underlying the formation of 2DEG at the STO heterointerface, particularly in LAO/STO heterostructures, is still under debate. Among the prevailing explanations [12-16], the electronic reconstruction across a sharp interface to minimize the electrostatic energy in the polar LAO film is the most favorable model [5, 16]. However, the recent discovery of electron doping by oxygen vacancies through unexpected redox reactions in STO substrates contributes often to the measured conductivity [17-19]. The reduction of the STO substrate arises mainly from its exposure to reactive species in the plasma plume present during film growth, particularly for films containing specific elements such as Al, Ti, Hf or Zr that have a large enthalpy of oxide formation [17, 19].

As far as the intensively-investigated LAO/STO interface is concerned, the LAO film is mostly grown by the physical vapor deposition technique of pulsed laser deposition (PLD). During film growth, the reaction occurs exclusively in the pressure range of $P_{O2} \leq 1 \times 10^{-3}$ mbar, where the plasma plume expands freely and shows negligible collisions with the background gas [20, 21]. In this range of gas pressure, the plasma plume species also exhibit higher kinetic energies, which allows layer-by-layer 2D film growth easier to achieve. However, it is the reactivity of the plume species rather than their high kinetic energy that results in the reduction of STO substrates, i.e. oxygen ions ($O^{2-}$) present in the STO substrate lattice provide extra oxygen sources for the film oxidization, in addition to oxygen from the target and the background oxygen gas [22]. At high temperatures, oxygen exchange between the film and substrate occurs faster and more extensively, therefore a significant reduction of STO during film deposition has been observed for oxygen pressure less than about $10^{-6}$ mbar [23,24].

For electron confinement at the LAO/STO interface, it is empirically accepted that the conduction observed in samples grown at low oxygen pressures (less than about $10^{-6}$ mbar) is dominated by oxygen vacancies and extends deeply into the substrate, whereas for samples grown at higher pressure (over about $10^{-5}$ mbar) the conduction is confined to the interface [9]. Generally, to prevent/suppress the oxygen-vacancies-related conduction, post annealing at high oxygen pressure is often adopted after the film deposition. However, it is known that the properties of STO substrates are very sensitive to the actual environmental conditions and history with respect to temperature and pressure [25]. Therefore, it has been reported that even minor differences in the actual procedures during heating and cooling may result in either insulating or metallic interfaces [14]. Remarkably, recent research shows that the introduction of a LaMnO3-based buffer layer [26, 27] or the substitution of Al by Cr [28] in LAO could strongly suppress the carrier density of the interfacial 2DEG as well as the reduction of STO during film deposition, without the need of post annealing. Importantly, this gives rise to an extreme enhancement in the electron mobility for the 2DEG fabricated at room temperature [26]. For example, the introduction of a single unit cell (uc) La$_{7/8}$Sr$_{1/8}$MnO$_3$ (LSMO) buffer layer at the interface of disordered-LaAlO$_3$/SrTiO$_3$ (d-LAO/STO) results in a mobility enhancement from typically 500 cm$^2$V$^{-1}$s$^{-1}$ to over 16000 cm$^2$V$^{-1}$s$^{-1}$ at 2 K, with a simultaneous decrease of the sheet carrier density to the order of $10^{12}$ cm$^{-2}$. This enables the clear observation of a quantum Hall effect at oxide interfaces [10]. However, the direct evidence for that the buffer layer suppressed the reduction of STO substrate, has not yet been investigated. Herein, we used $^{18}$O isotope exchanged SrTi$^{18}$O$_3$ as substrates to create 2DEG at room temperature with and without the LSMO buffer layer. By mapping the oxygen profile across the interface, we provided unambiguous evidence on the reduction of STO during the formation of the 2DEG without the LSMO buffer layer as well as a clear trend of suppressed reduction of STO upon the introduction of the manganite buffer layer.

The metallic interfaces, i.e. 2DEGs, were fabricated by depositing LAO or yttria-stabilized zirconia (YSZ) films on $O^{18}$-exchanged STO substrates or $O^{18}$-exchanged LSMO-buffered STO substrates by PLD at room temperature, where the films are in the form of amorphous or disordered structure (d-LAO and d-YSZ, respectively, for the $LaAlO_3$ and YSZ films). The adopted room temperature procedure is mainly applied to alleviate the thermal activated oxygen exchange between the substrate and the film. Therefore, any changes in the $O^{18}$ depth profile are induced by the deposited film itself. It is notable that, although extensive oxygen exchange between STO and LAO has been determined by secondary ion mass spectrometry (SIMS) depth profiling for samples grown at high temperatures [22], no oxygen exchange between labeled STO and LAO grown at room temperature was detected [22]. Instead of using SIMS, we here used elastic recoil detection analysis (ERDA) to analyze the oxygen isotopes of the samples, which offers the possibility of absolute calibration of the oxygen depth profile. The $O^{18}$-exchanged substrates were obtained by sintering bare commercial STO or LSMO-buffered STO substrates at 1100 °C for 48 h. This process led to an $O^{18}$ exchange of close to 70% for the bare substrates while the average $^{18}O$ content of the buffered substrates was only between 35 and 40%. Single-crystalline LAO and YSZ were used as targets for the deposition of d-LAO and d-YSZ films, respectively. The distance between the target and the substrate was 5.5 cm. A KrF (λ=248 nm) laser with a repetition rate of 1 Hz and laser fluence of 4.0 J/cm$^2$ was applied. During film growth, the oxygen background pressure was kept at $1\times10^{-4}$ mbar. The film thickness in addition to the composition of d-LAO and d-YSZ films was determined by Rutherford backscattering spectrometry (RBS). Electrical characterization was performed using a 4-probe Van der Pauw method with ultrasonically wire-bonded aluminum wires as electrodes. For the ERDA analysis, a 13 MeV $^{127}I$ beam was used under 18° incidence angle.

Fig.1(a) illustrates the sketch of the LSMO-buffered oxide interfaces. Based on d-LAO/STO$^{16}$ and d-YSZ/STO$^{16}$ heterostructures without $O^{18}$-exchanged substrates, the LSMO buffer layer is optimized to be 1 uc and 5 uc in thickness, respectively [29]. Figs. 1 (b) and (c) show the typical experimental RBS spectra using 2 MeV He$^+$ for d-LAO/STO$^{18}$ and d-YSZ/STO$^{18}$ along with optimized simulations to obtain the film composition. RBS can provide information on the composition, structure and thickness of epitaxial films. Herein RBS is mainly used to determine the film thickness, which gives rise to 44 nm and 45 nm for the d-LAO/STO$^{18}$ and d-YSZ/STO$^{18}$ without buffer layer, respectively. For the buffered samples, the film thickness is determined to be slightly thinner, which are 32 nm for d-LAO/LSMO/STO$^{18}$, and 36 nm for d-YSZ/LSMO/STO$^{18}$. Fortunately, this difference will not influence the transport properties and the oxygen depth profile as discussed in the following. Moreover, we detected Hf contaminants for the d-YSZ/STO heterostructures (Fig. 1c). Since such contamination is absent for the d-LAO/STO samples (Fig.1a), we assume the contamination comes from the impurity of the single-crystalline YSZ target used for the film deposition, which is common for zirconia-based compounds [30].

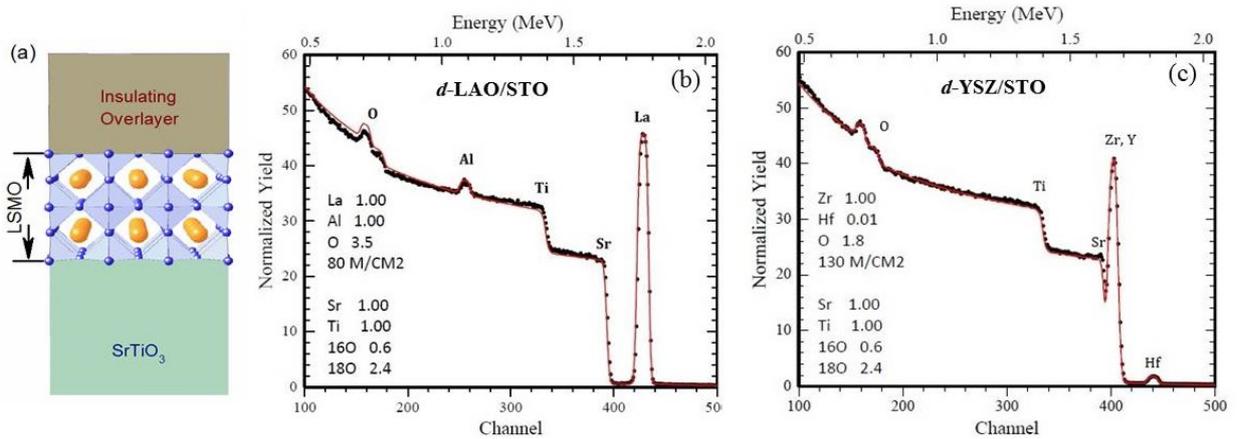

**Fig.1** Sketch of the buffered heterostructure (a) and the typical experimental RBS spectra using 2 MeV He$^+$ for $d$-LAO/STO$^{18}$ (b) and $d$-YSZ/STO$^{18}$ (c) along with optimized simulations to obtain the film composition (red lines).

As reported previously, both $d$-LAO/STO and $d$-YSZ/STO heterostructures are metallic although each of the individual components is highly insulating [17]. With the O$^{18}$-exchanged STO$^{18}$ substrates, both heterointerfaces remain metallic. However, compared to the $d$-LAO/STO$^{16}$ and $d$-YSZ/STO$^{16}$ on conventional STO$^{16}$ substrates, the sheet resistance, $R_s$, of both $d$-LAO/STO$^{18}$ and d-YSZ/STO$^{18}$ shows a remarkable upturn at $T\sim23$ K, as shown in Fig.2 (a). This is a typical characteristic related to the ferroelectricity of STO$^{18}$ induced by the oxygen isotope exchange, which also indicates that more than 35% of the oxygen of the STO substrate is substituted with the isotope O$^{18}$ [31]. The latter is well confirmed by the ERDA measurements (Fig. 3). Upon introducing the LSMO buffer layer, where the thickness is 1 uc and 5 uc for $d$-LAO/LSMO/STO$^{18}$ and $d$-YSZ/LSMO/STO$^{18}$ respectively, the most remarkable change lies at the suppression in the electron carrier density, $n_s$ at room temperature [Fig. 2(b)]. Specifically, for $d$-LAO/STO$^{18}$, $n_s$ is decreased from $8.9\times10^{13}$ cm$^{-2}$ for non-buffered sample to $2.5\times10^{13}$ cm$^{-2}$ upon introducing the 1 uc LSMO buffer layer, where the carrier freezing out phenomenon upon cooling (below 100 K) becomes also negligible. For $d$-YSZ/STO$^{18}$, $n_s$ is decreased from $1.4\times10^{14}$ cm$^{-2}$ for non-buffered sample to $8.5\times10^{13}$ cm$^{-2}$ upon introducing the LSMO buffer layer. But the carrier freezing out phenomenon remains here, indicative of the presence of charge-trap defect states even in the 5uc-LSMO buffered $d$-YSZ/STO$^{18}$. Notably, a large enhancement in the electron mobility from 385 cm$^2$V$^{-1}$s$^{-1}$ to 2235 cm$^2$V$^{-1}$s$^{-1}$ is achieved for the $d$-LAO/LSMO/STO$^{18}$ system at 2 K (Fig.2c), although this magnitude of mobility enhancement is relatively small compared to the LSMO-buffered $d$-LAO/STO$^{16}$ [26,27]. The relatively low electron mobility for the best $d$-LAO/LSMO/STO$^{18}$ sample could result from the scattering related to the ferroelectric instability induced by the oxygen isotope exchange [31]. It should be also noted that the $d$-LAO/STO$^{18}$ with and without the LSMO buffer layer exhibit comparable sheet carrier densities at 2 K, therefore the higher mobility for buffered samples should imply a lower level of defects/impurities, rather than other effects such as electron-electron interactions.

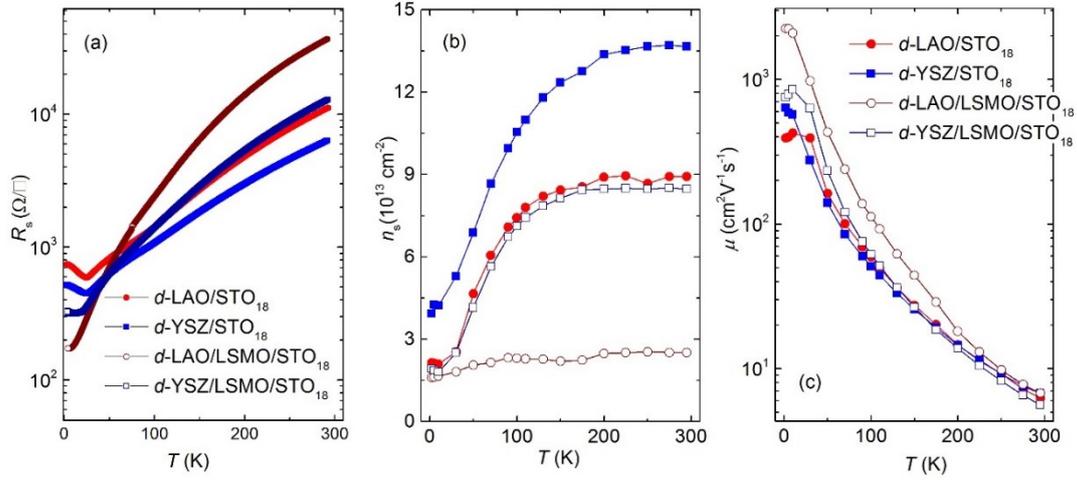

**Fig.2** The temperature-dependent transport properties (a) $R_s$; (b) $n_s$; and (c) $\mu$ for both $d$-LAO/STO[18] and d-YSZ/STO[18] heterointerfaces with and without LSMO buffer layers.

It has been revealed, by in-situ X-ray photoelectron spectroscopy (XPS) measurements, that a high concentration of $Ti^{3+}$ is present in $d$-LAO/STO without the buffer layer [17]. The amount of $Ti^{3+}$ deduced from the XPS measurement [17] is found to be much larger than the mobile electrons accounting for the conduction, which indicates the formation of oxygen vacancies in STO substrates. In contrast, all LSMO-buffered $d$-LAO/STO samples show no discernible $Ti^{3+}$ signal, implying a strongly suppressed $Ti^{3+}$ or redox reaction far below the detection limit of XPS measurements. Meanwhile, the LSMO layer shows a remarkable shake-up satellite peak, indicative of the presence of $Mn^{2+}$ in the buffer layer [26]. Recent non-destructive resonant x-ray reflectometry experiments reveal that, the $Mn^{2+}$ in $d$-LAO/LSMO/STO is in a combination of octahedral ($O_h$) and tetrahedral coordinations ($T_d$), indicative of the formation of oxygen vacancies in the buffer layer. More remarkably, for the $d$-YSZ/LSMO/STO with 5 uc LSMO, the top surface of the perovskite LSMO is transformed to a brownmillerite structure containing divalent Mn upon overlayer deposition, i.e. the LSMO buffer layer consists of a section of a perovskite LSMO and a brownmillerite like section at the interface to the $d$-YSZ [29]. This strongly suggests that redox reactions, if happening, are strongly confined within the manganite buffer layer, while the STO substrate is protected from reduction upon the introduction of the LSMO buffer layer.

The occurrence of interfacial redox reaction and its difference between the buffered and unbuffered heterostructures should be detectable from the ratio of $^{16}O$ and $^{18}O$ as a function of depth. Fig.3 shows the oxygen depth profile (both $^{18}O$ and $^{16}O$) for both LSMO-buffered and non-buffered $d$-LAO/STO[18] and $d$-YSZ/STO[18] heterointerfaces. Note that the depth profile of the total oxygen concentration for the unbuffered and buffered $d$-LAO/STO interfaces (Figs. 3 a and c, respectively), indicates a kink in proximity to the interface. However, since the material composition drastically changes at the interface, it remains unclear whether such a kink is evidence of the reduction of the STO substrate or artefact due to the abrupt change in energy loss of passing ions at the interface. Moreover, there will be always $O^{16}$ exchange between the sample and the environment (such as $O^{16}$ uptaking as indicated by the conductivity aging effect). Therefore, in the following, the comparison between the buffered and unbuffered samples is mainly focused on the $O^{18}$ depth profile. To allow a comparison between buffered and unbuffered samples with different film thicknesses, the depth has been rescaled in a way that the substrate edges seen in ERDA correspond to the layer thickness measured by RBS. As shown in Figs.3 (a) and (b), for non-buffered $d$-LAO/STO[18] and $d$-YSZ/STO[18] heterointerfaces, respectively, the outward diffusion of $O^{18}$ from the STO substrate is indeed observed unambiguously upon the film deposition. Notably, the two types of the heterostructures give rise to almost the same depth profile. When comparing the nonbuffered samples (Figs. 3a and b) to the buffered samples (Figs. 3c and d), two interesting features can be determined: Firstly, the $^{18}$O content of the whole buffered STO substrate is much lower than that of the unbuffered sample, approximately 40 at.% vs. 80 at.% as shown in Figs. 3 (a) and (c) for buffered and unbuffered $d$-LAO/STO$^{18}$, respectively. This is unexpected since the buffered and unbuffered STO substrates experienced exactly same $^{18}$O exchange procedure. Secondly, due to the limited depth resolution, it is hard to determine whether there is a difference in the $^{18}$O deficient region on the STO side in proximity to the interface between the buffered and unbuffered samples. However, it is likely that larger outward diffusion of O$^{18}$ into the top films is observed for the unbuffered sample (Figs.3a and b). This is also supported by the fact that there is certain $^{18}$O present at the surface of the unbuffered samples whereas there is no significant $^{18}$O signal at the surface of buffered samples although these films are thinner. This finding suggests less reduction of the STO substrate in the buffered sample, as expected from the transport properties shown in Fig.2. Note that, in order to compare precisely the oxygen depth profile between the buffered and unbuffered samples, the $^{18}$O concentrations of STO$^{18}$ substrates are normalized to the same level as shown in Fig. 3a and b), where the depth profiles of buffered samples are reproduced based on the same diffusion parameter revealed in Figs. 3 (c) and (d).

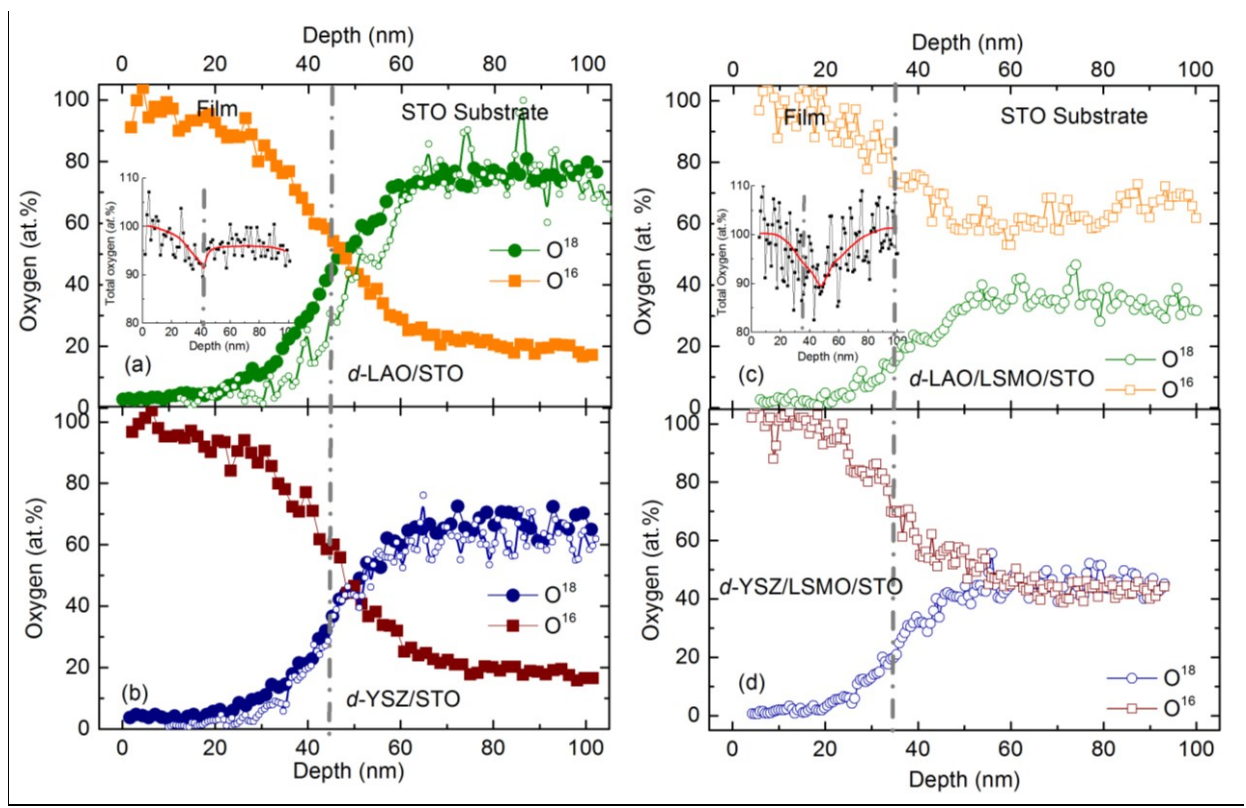

**Fig.3** The oxygen isotope concentration depth profile for both $d$-LAO/STO$^{18}$ (a, c) and $d$-YSZ/STO$^{18}$, (b, d) heterointerfaces with and without LSMO buffer layers determined by ERDA. The insets of a and c show the corresponding depth profile of the total oxygen concentration. The concentration profiles are normalized to $^{18}$O + $^{16}$O = 100 % at the surface. This normalization slightly fails in the substrate due to the drastic composition change of the material.
The depth scale is only approximate but the position of the interface to the substrate is used for a correct calibration. The grey dashed lines indicate the interface position. For comparison, modified $^{18}$O depth profiles of buffered samples, based on the same diffusion parameter revealed in (c) and (d), respectively, were also shown in (a) and (b).

The overall smaller $^{18}$O content (~40 *at.*%) in the LSMO buffered STO substrate compared to the unbuffered STO (~80 *at.*%)) is unusual. This indicates that the LSMO buffer layer which was epitaxially-grown in a layer-by-layer mode suppressed the oxygen exchange of the LSMO/STO heterostructures even at high temperatures. This could be due to the fact that the oxygen exchange kinetic parameters of epitaxial layers of LSMO resemble those of grains in polycrystalline LSMO, whose diffusion and surface exchange values are found to be approximately two to three orders of magnitude lower than those of the grain boundaries [32]. The less reduction of the STO substrate in buffered samples (Figs.3c and d) could also arise from the suppressed oxygen exchange kinetic parameters upon introducing the LSMO buffer layer. But this can hardly explain the fact that only 1 uc LSMO buffer layer could block strongly the redox reaction at the *d*-LAO/STO interface as indicated from the transport properties (the temperature-independent $n_s$ as shown in Fig.2b) as well as previous XPS measurements [26]. Therefore, we believe that the electronic structure of buffered samples also plays an important role: the LSMO buffer layer has an empty or partially filled $e_g$ subband which is lower than the Ti 3*d* $t_{2g}$ band in STO [33,34]. In this vein, the suppressed $n_s$ of the buffered samples as well as the suppressed redox reaction could be well explained within the charge transfer induced modulation doping scheme as reported in Ref.26. Electrons, donated from the top *d*-LAO layer by electronic reconstruction or by redox reactions during the film deposition, should first fill the empty or partially filled subband of the LSMO layer before filling the well at the interface between the STO and the LSMO. Therefore, the redox reaction can be mainly confined in the LSMO buffer layer as revealed by the resonant x-ray reflectometry experiments [29]. Unfortunately, such confinement cannot be determined by the isotope concentration depth profile since it is beyond the depth resolution limit.

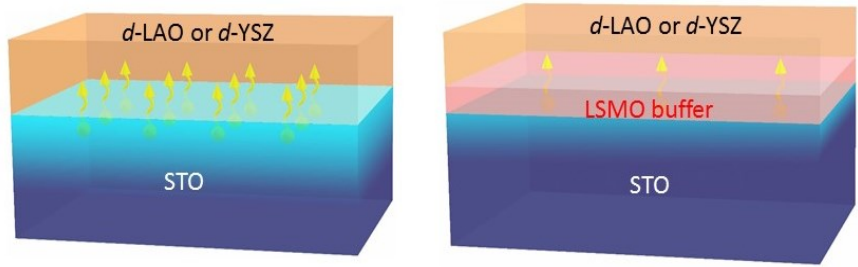

**Fig.4.** Sketch of the oxygen outward diffusion without (a) and with (b) the LSMO buffer layer. The introduction of the LSMO buffer layer suppresses the oxygen outward diffusion thus reduction in the STO substrate. The arrows indicate the outward diffusion of lattice oxygen.

In summary, we used oxygen isotope tracing to investigate the redox reaction of STO 2DEG heterostructures with and without a LSMO-buffer. We have found that the LSMO buffer layer suppresses strongly the carrier density of the metallic interface of *d*-LAO/STO[18] and *d*-YSZ/STO[18] heterointerfaces. For unbuffered samples, as illustrated in Fig.4a, significant reduction into the STO substrate as proposed previously [17] is confirmed. For buffered samples, as illustrated in Fig.4b, the oxygen outward diffusion as observed in unbuffered samples is suppressed, which indicates less reduction of the STO substrate during the formation of 2DEG. This could result in lower impurity scattering in the buffered oxide interface and explain the enhanced electron mobility.

**Acknowledgement:**

We acknowledge the technical help from J. Geyti, and the financial support from the Innovation Fund Denmark.